\documentclass[dvipdfmx]{pasj01}
\draft
\usepackage{color}
\usepackage[normalem]{ulem}

\usepackage{multirow}
\usepackage[]{graphicx}

\begin{document} 

\title{The formation of the H{\sc ii} regions N83 and N84 in the Small Magellanic Cloud triggered by colliding H{\sc i} flows}
\author{Takahiro \textsc{OHNO}\altaffilmark{1}}
\author{Yasuo \textsc{FUKUI}\altaffilmark{1,2}}
\author{Kisetsu \textsc{TSUGE}\altaffilmark{1}}
\author{Hidetoshi \textsc{SANO}\altaffilmark{3}}
\author{Kengo \textsc{TACHIHARA}\altaffilmark{1}}
\email{t.ohno@a.phys.nagoya-u.ac.jp}
\altaffiltext{1}{Department of Physics, Nagoya University, Furo-cho, Chikusa-ku, Nagoya 464-8601, Japan}
\altaffiltext{2}{Institute for Advanced Research, Nagoya University, Furo-cho, Chikusa-ku, Nagoya 464-8601, Japan}
\altaffiltext{3}{National Astronomical Observatory of Japan, Mitaka, Tokyo 181-8588, Japan}


\maketitle

\begin{abstract}
LHA 115-N 83 (N83) and LHA 115-N 84 (N84) are H{\sc ii} regions associated with the early stage of star formation located in the Small Magellanic Cloud (SMC). We have analyzed the new H{\sc i} data taken with the Galactic Australian Square Kilometre Array Pathfinder survey project at a high angular resolution of 30\arcsec. We found that the two clouds, having $\sim$40 km s$^{-1}$ velocity separation, show complementary distribution with each other, and part of the H{\sc i} gas is dispersed by the ionization. In addition, the Atacama Large Millimeter/submillimeter Array observations revealed clumpy CO clouds of 10$^{5}$ $M_{\odot}$ in total over an extent of 100 pc, which are also well correlated with the H{\sc ii} regions. There is a hint of displacement between the two complementary components, which indicate that the red-shifted H{\sc i} cloud is moving from the north to the south by $\sim$100 pc. This motion is similar to what is found in NGC~602 \citep{2020arXiv200513750F}, suggesting a large scale systematic gas flow. We frame a scenario that the two components collided with each other and triggered the formation of N83, N84, and six O-type stars around them in a time scale of a few Myr ($\sim$60 pc / 40 km s$^{-1}$). The supersonic motion compressed the H{\sc i} gas to form the CO clouds in the red-shifted H{\sc i} cloud, some of which are forming O-type stars ionizing the H{\sc ii} regions in the last Myr. The red-shifted H{\sc i} cloud probably flows to the direction of the Magellanic Bridge. The velocity field originated by the close encounter of the SMC with the Large Magellanic Cloud 200 Myr ago as proposed by \citet{1990PASJ...42..505F}.
\end{abstract}

\section{Introduction} \label{sec:intro}
Dynamical interactions between galaxies are a common process in the Universe and exchange of the interstellar medium (ISM) between galaxies is happening often to drive the galaxy evolution. It is also likely that the gas motion produces shocks to trigger star formation. There are a number of interacting galaxies as a most typical case, the Antennae Galaxies where many young massive clusters are probably formed by the interactions (\cite{1998Natur.395..859G,2000ApJ...542..120W,2014ApJ...795..156W,2019ApJ...874..120F,2019arXiv190905240T}, \yearcite{2020arXiv200504075T}).

The Magellanic Clouds are a tidally interacting system with close encounters in the past. \citet{1990PASJ...42..505F} proposed that the most recent encounter 200 Myrs ago produced H{\sc i} flows driven by the tidal force between the Large Magellanic Cloud (LMC) and the Small Magellanic Cloud (SMC). \citet{2017PASJ...69L...5F} analyzed the H{\sc i} data in the LMC taken with the Australia Telescope Compact Array (ATCA) / Parkes telescope and discovered signatures of collision of H{\sc i} gas having two velocity components with 60 km s$^{-1}$ velocity separation, the L and D components (\cite{1991IAUS..148...63R}, \yearcite{1993LNP...416..115R}; \cite{1992A&A...263...41L}). The collision signatures are the complementary distribution between them and the bridge features connecting the two components in velocity. It is suggested that the cluster RMC136 (R136), which is most massive in the Local Group, and 400 OB stars in the H{\sc i} ridge of the LMC were formed by the kpc-scale trigger in the last 10 Myr. Detailed simulations by \citet{2007PASA...24...21B} reproduced the H{\sc i} ridge and offer supporting theoretical results of the interaction. The results by \citet{2017PASJ...69L...5F} indicate that a H{\sc i} collision, not a CO collision, can trigger star formation contrary to the usual picture of star formation from molecular gas. Because the galactic interaction in the Magellanic system is a unique event very close to the sun, the consequences of the interaction require thorough investigation in order to understand the role of H{\sc i} collision in star formation. It is suggested that the close encounter 200 Myr ago also drove the H{\sc i} flow which is being driven from the SMC to the LMC \citep{2007MNRAS.381L..11M}. The SMC is known as an active galaxy in star formation as well as the LMC (e.g., \cite{2001PASJ...53L..45M}). It is therefore important to explore a possibility that the tidal interaction triggers star formation in the SMC in an effort to comprehend the whole picture of star formation in the Magellanic system.

In NGC~602 which is located in the extreme southeast wing of the SMC, the formation of NGC~602 and nearby ten O-type stars triggered by H{\sc i} gas flows are revealed by \citet{2020arXiv200513750F}. The present work is a next step to pursue the star formation in the SMC by analyzing the high-resolution H{\sc i} data taken by the Galactic Australian Square Kilometre Array Pathfinder (GASKAP). We here focus on the H{\sc ii} regions including LHA 115-N 83 (N83) and LHA 115-N 84 (N84) associated with the early stage of star formation \citep{1956ApJS....2..315H}. The paper is organized as follows. Section \ref{sec:obs} give descriptions on the H{\sc i} and $^{12}$CO($J$ = 2--1) datasets, and Section \ref{sec:results} explains the results of the analysis. In Section \ref{sec:discuss} we discuss the implications of the results of the H{\sc i} gas disruption and its relationship to the star formation.

\section{Observations}\label{sec:obs}
\subsection{H{\sc i}}
The H{\sc i} observations were made as part of the test run GASKAP of Australian Square Kilometre Array Pathfinder (ASKAP) which aims at surveying $\pm10{^\circ}$ of the galactic plane and the Magellanic Clouds \citep{2009IEEEP..97.1507D,2018NatAs...2..901M}. The angular resolution is 35\arcsec~$\times$ 27\arcsec~(corresponds to 10 pc at the distance of the SMC) with a position angle of $89\fdg62$ and rms noise level is 0.91 K in 3.9 km s$^{-1}$ velocity resolution. 

\subsection{$^{12}$CO($J$ = 2--1)}
To derive physical properties of CO clouds surrounding N83 and N84, we used archival CO data sets obtained with the Atacama Large Millimeter/submillimeter Array (ALMA) Band 6 (211--275 GHz) as Cycle 6 project (\#2018.1.01319.S, PI: L. Johnson). Observations of $^{12}$CO($J$ = 2--1) line emission at 230.403 GHz was conducted in 2018 November using 3--4 antennas of a total power (TP) array. The OTF mapping mode with Nyquist sampling was used. The map size is about a 12\arcmin~$\times$ 10\arcmin~rectangular region centered at ($\alpha_\mathrm{J2000}$ = $1^{\mathrm{h}}14^{\mathrm{m}}18\fs157$, $\delta_\mathrm{J2000}$ = $-73{^\circ}17\arcmin4\farcs545$). We utilized the product data set through the pipeline processes using the Common Astronomy Software Application (CASA; \cite{2007ASPC..376..127M}) package version 5.4.0-68 with the Pipeline version 42030M (Pipeline-CASA54-P1-B). The beam size is $\sim$29\arcsec~for $^{12}$CO($J$ = 2--1), corresponding to a spatial resolution of $\sim$9 pc at the distance of the SMC. The typical noise fluctuations of the $^{12}$CO($J$ = 2--1) is $\sim$0.025 K for velocity resolution of 0.32 km s$^{-1}$.

\section{Results}\label{sec:results}
\subsection{The spatial distribution of H{\sc i} toward whole the SMC}
Figure \ref{fig1} shows the integrated intensity map of ASKAP H{\sc i} toward whole the SMC \citep{2009IEEEP..97.1507D,2018NatAs...2..901M}. The H{\sc i} clouds are extended on a few kpc scales. The SMC consists of two H{\sc i} components. One of them spatially extends from northeast to southwest (hereafter, main body). The other extends southeast of the SMC near $\delta_\mathrm{J2000} \sim -73{^\circ}$ (hereafter, wing). The wing is thought to be connected to the Magellanic Bridge. The N83/N84 region is located toward the southeastern part of the wing. Almost all O-type stars are located in the main body and toward the N83/N84 region. The white dashed box including N83 and N84 in Figure \ref{fig1} shows a focused region in the present paper (Figures \ref{fig2}--\ref{fig6}).

\subsection{The galactic rotation velocity component of the SMC}
According to \citet{2004ApJ...604..176S} and \citet{2019MNRAS.483..392D}, the SMC has the galactic rotation velocity component. In order to avoid confusion in the analysis, we subtract the galactic rotation velocity component by shifting the H{\sc i} spectra in the velocity direction. We define radial velocity of the SMC $V_{\rm {offset}}$ as follows;
\begin{equation}
V_{\rm {offset}} = V_{\rm {LSR}} - V_{\rm {rotation}},
\end{equation}
where $V_{\rm {rotation}}$ is the rotation velocity of the SMC. See Appendix for more details. Two main velocity components are found toward the N83/N84 region as $V_{\rm {offset}}$ = $-$50.0--$-$29.9 km s$^{-1}$ and $V_{\rm {offset}}$ = $-$19.9--13.0 km s$^{-1}$, respectively (see Figures \ref{fig2} and \ref{fig4}).

\subsection{Two main velocity components toward the N83/N84 region}
Figure \ref{fig2} shows the typical H{\sc i} spectra derived from two positions around the N83/N84 region. Figures \ref{fig2}(a) and \ref{fig2}(b) show the H{\sc i} spectra at position A ($\alpha_\mathrm{J2000}$ = $1^{\mathrm{h}}23^{\mathrm{m}}11\fs5$, $\delta_\mathrm{J2000}$ = $-73{^\circ}36\arcmin40\farcs2$) and position B ($\alpha_\mathrm{J2000}$ = $1^{\mathrm{h}}10^{\mathrm{m}}19\fs6$, $\delta_\mathrm{J2000}$ = $-73{^\circ}17\arcmin57\farcs2$), respectively. The velocity range is defined by visual inspection using the right ascension-velocity diagram (Figure \ref{fig4}) and H{\sc i} profiles; $V_{\rm {offset}}$ = $-$50.0--$-$29.9 km s$^{-1}$ for position A (hereafter, the blue cloud) and $V_{\rm {offset}}$ = $-$19.9--13.0 km s$^{-1}$ for position B (hereafter, red cloud). The peak velocity is also derived from a Gaussian fitting; $V_{\rm {offset}}$ = $-$37.5 km s$^{-1}$ for position the blue cloud and $V_{\rm {offset}}$ = $-$2.4 km s$^{-1}$ for the red cloud. Figure \ref{fig2}(c) shows the map of moment 1 toward the N83/N84 region. The blue cloud is distributed on the eastern part of the region, and the red cloud is distributed on the western part. It is noteworthy that N83A, N83B, N83C, N84A, N84B, N84C, and N84D are located in the red cloud.

We compare the spatial distribution of the blue cloud in Figure \ref{fig3}(a) with that of the red cloud in Figure \ref{fig3}(b). The blue cloud has a cavity with a diameter $\sim$20\arcmin, which is centered at ($\alpha_\mathrm{J2000} \sim 1^{\mathrm{h}}12^{\mathrm{m}}$, $\delta_\mathrm{J2000} \sim -73{^\circ}15\arcmin$) including N83, N84, and other six O-type stars (blue crosses in Figure \ref{fig3}; \cite{2010AJ....140..416B}). On the other hand, the distribution of the red cloud is relatively concentrated around N83 and N84, corresponding to the cavity of the blue cloud.

We confirmed that the blue and red clouds are in the right ascension-velocity diagram in Figure \ref{fig4}. The blue and red horizontal lines show the ranges of the blue and red clouds, respectively. We see the blue and red clouds are clearly separated each other. We find connecting component between the blue and red clouds at $\alpha_\mathrm{J2000}$ = $1^{\mathrm{h}}10^{\mathrm{m}}$--$1^{\mathrm{h}}15^{\mathrm{m}}$ and $\alpha_\mathrm{J2000}$ = $1^{\mathrm{h}}22^{\mathrm{m}}$--$1^{\mathrm{h}}27^{\mathrm{m}}$ (hereafter, the bridge component).

We summarize the physical properties of the blue cloud, the red cloud, and the bridge component in Table \ref{t1}. We derived column density $N$(H{\sc i}), assuming optically thin, as follows;
\begin{equation}
N(\rm {H\textsc{i}}) = 1.8224 \times 10^{18} \int_{\it v_{\rm {1}}}^{\it v_{\rm {2}}} \it {\Delta T_{\rm {b}} dv}~[\rm {cm^{-2}}],
\end{equation}
where $v_1$ and $v_2$ represent the integration velocity range, and $T_{\rm {b}}$ represents the brightness temperature of H{\sc i}. We define cloud radius $R_{\rm {HI}}$ as an effective radius derived by using the following equation;
\begin{equation}
R_{\rm {HI}} = (A / \pi)^{0.5},
\end{equation}
where $A$ is the area of the region surrounded by a contour of 20\% of the peak integrated intensity. The peak velocities of the blue and red clouds are $V_{\rm {offset}}$ = $-$37.5 km s$^{-1}$ and $V_{\rm {offset}}$ = $-$2.4 km s$^{-1}$, respectively. This suggests that the velocity separation of the two clouds is 35.1 km s$^{-1}$. The peak column density of each cloud is $\sim$1.2--4.9 $\times$ 10$^{21}$ cm$^{-2}$. Masses of the blue and red clouds are roughly the same ($\sim$1.0 $\times$ 10$^{7}$ $M_{\odot}$), while mass of the bridge component is $\sim$0.3 $\times$ 10$^{7}$ $M_{\odot}$. Typical cloud radius of each cloud is $\sim$400--600 pc.

Figure \ref{fig5} shows an overlay map of the blue and red clouds. We find that the blue and red clouds show complementary spatial distribution; the cavity in the blue cloud shows a good spatial correspondence with the denser part of the red cloud in Figure \ref{fig5}(a). We find a displacement of 57 pc between the blue and red clouds in Figure \ref{fig5}(b). In order to derive the displacement, we applied an algorithm developed by \citet{2020arXiv200313925F}. Figure \ref{fig6} shows the distribution of the bridge component superposed on contours of the blue and red clouds. The bridge component is distributed in the cavity of the blue cloud. In particular, it is relatively dense at ($\alpha_\mathrm{J2000}$ = $1^{\mathrm{h}}15^{\mathrm{m}}$, $\delta_\mathrm{J2000}$ = $-73{^\circ}06\arcmin$), corresponding to boundaries of the blue and red clouds.

In order to analyze the physical relation between H$\alpha$ and H{\sc i} cloud, we focus on the H{\sc ii} region N83A. Figure \ref{fig7}(a) shows the H{\sc i} contour map integrated in the velocity range of $V_{\rm {offset}}$ = $-$10.0--0.0 km s$^{-1}$ superposed on the optical image toward N83A obtained from DSS2. The H{\sc i} cloud has an intensity depression toward N83A, possibly suggesting that the intensity depression was formed by stellar feedback, and is physically related to N83A. We also show the right ascension-velocity diagram in Figure \ref{fig7}(b). H{\sc i} cloud has a cavity at $\alpha_\mathrm{J2000}$ = $1^{\mathrm{h}}13^{\mathrm{m}}45^{\mathrm{s}}$--$1^{\mathrm{h}}13^{\mathrm{m}}54^{\mathrm{s}}$, whose diameter is $\sim$15 pc in the right ascension.

\subsection{The CO clouds toward the N83/N84 region}
Figure \ref{fig8} shows the $^{12}$CO($J$ = 2--1) contour map superposed on the optical image of N83 and N84 obtained from DSS2. We identified 29 CO clouds of A1--Y using ALMA data shown in Figure \ref{fig8} according to criteria as follows. We also derived the physical properties of the CO clouds from a Gaussian fitting. 
\begin{enumerate}
\item An area enclosed by the contour of 0.374 K km s$^{-1}$ (equal to 5$\sigma$ level for the integration for the velocity range of $V_{\rm {LSR}}$ = 148.2--175.2 km s$^{-1}$) in Figure \ref{fig8} is defined as an individual cloud.
\item When the major axis of an area is smaller than the beam size (29\arcsec), the area is not defined as an individual cloud. 
\item When there are multiplex peaks in the same area, the peaks are divided by the minimum in the intensity map and are defined as independent.
\item When a CO spectrum has two peak profiles, we conducted the following operation.
\begin{enumerate}
\item We define each peak as an independent cloud in the case for which the velocity difference between the two peaks is larger than 4.2 km s$^{-1}$ (equal to 2 times the typical line width) (A1, A2, C1, C2, S1, S2, W1, and W2).
\item We apply two gaussian fitting in the case for which the velocity difference between the two peaks is smaller than 4.2 km s$^{-1}$ (equal to 2 times the typical line width) (L and T).
\end{enumerate}
\end{enumerate}
The CO-derived mass $M_{\rm {CO}}$ is obtained from the following equations;
\begin{equation}
M_{\rm {CO}} = m_{\rm H} \mu \sum_{i} [D^2 \Omega N_i(\rm {H_2})],
\end{equation}
\begin{equation}
N({\rm {H_2}}) = X_{\rm {CO}} W(\rm {CO})~[\rm {cm^{-2}}],
\end{equation}
where $m_{\rm H}$ is the mass of atomic hydrogen, $\mu$ is the mean molecular weight relative to atomic hydrogen, $D$ is the distance to the source in units of cm, $\Omega$ is the solid angle subtended by a unit grid, $N_i(\rm {H_2})$ is the molecular hydrogen column density for each pixel in units of cm$^{-2}$, $X_{\rm {CO}}$ is CO-to-H$_2$ conversion factor, and $W$(CO) is the $^{12}$CO($J$ = 1--0) integrated intensity. We applied equation (4) for the area of the region surrounded by a contour of 50\% of the peak integrated intensity. We adopted $\mu$ = 2.7 and $X_{\rm CO}$ = 7.5 $\times$ 10$^{20}$ cm$^{-2}$ (K km s$^{-1}$)$^{-1}$ \citep{2017ApJ...844...98M}. We derived $W$(CO) from our $^{12}$CO($J$ = 2--1) data using $^{12}$CO($J$ = 2--1)/$^{12}$CO($J$ = 1--0) ratio of 0.9 \citep{2003ApJ...595..167B}. We summarize the physical properties of CO clouds in Table \ref{t2}. The peak velocities of CO clouds derived from a Gaussian fitting are $V_{\rm {LSR}}$ = 151.3--170.1 km s$^{-1}$ and the total CO-derived mass is $\sim$10$^5$ $M_{\odot}$. We derived cloud size $D_{\rm {CO}}$ as follows;
\begin{equation}
D_{\rm {CO}} = 2 \times (A / \pi)^{0.5},
\end{equation}
where $A$ is the area of the region surrounded by a contour of 50\% of the peak integrated intensity.

\section{Discussion} \label{sec:discuss}
\subsection{Star formation in N83 and N84} \label{sec:discuss1}
The SMC is very active in high-mass star formation along with the LMC. It has been a puzzle how the active star formation is taking place in the low density outer part, the southeast wing, in the SMC. There are two outstanding star formation spots, the N83/N84 region and the NGC~602 region. \citet{2020arXiv200513750F} presented a scenario that H{\sc i} colliding flows triggered the formation of NGC~602 and eleven O-type stars in the interface layer of the collision in the last several Myr.

Star formation in N83 and N84 was intensively studied by several authors \citep{1987A&A...178...25T,1988A&A...194...11L,2004A&A...423..919B}. These authors showed that the age of N83 and N84 is 2.5--5 Myr and even older B stars with age of 12--30 Myr are distributed in the central part. \citet{2004A&A...423..919B} argued that sequential star formation happened in part of a supergiant shell in a timescale of 10 Myr, and suggested that a supernova took place 30 Myr ago. Another view is presented by \citet{2003ApJ...595..167B} that the age of the shell is as young as $\sim$2.3 Myr based on the molecular shell of 25 pc radius expanding at 6.5 km s$^{-1}$ by adopting the interstellar bubble model of \citet{1977ApJ...218..377W}. In summary, the region possibly includes objects of an age spread of less than 2 Myr to more than 10 Myr. The present ALMA data revealed the H{\sc ii} regions are associated with 29 CO clouds, most of which are not associated with star formation, suggesting that star formation will be continuing in future.

\subsection{Cloud-cloud collision in southeast of the SMC} \label{sec:discuss2}
We have analyzed the new H{\sc i} data taken by the GASKAP at 30\arcsec~resolution toward the H{\sc ii} regions N83 and N84 over 1 kpc $\times$ 1 kpc. We found that there are two velocity components peaked at $V_{\rm {offset}}$ = $-$37.5 km s$^{-1}$ and $V_{\rm {offset}}$ = $-$2.4 km s$^{-1}$ in the region, which show complementary distribution with a displacement of 57 pc as well as the bridge feature connecting them in velocity. 

Based on the results, we present a scenario that the two components collided with each other. In the scenario, the collisional compression triggered the formation of N83, N84 and six O-type stars in the compressed region mainly in the southern edge of the red cloud. The total stellar mass of the region is estimated to be $\sim$120 $M_{\odot}$ (= 6 $\times$ 20 $M_{\odot}$, if a single stellar mass is assumed to be 20 $M_{\odot}$). In the scenario, the compression took place in the region of a 100 pc diameter which corresponds to the extent of the 29 CO clouds. It seems likely that the clumpy CO clouds of $\sim$10$^{5}$ $M_{\odot}$ were formed by the compression from the H{\sc i} gas, with column density 3 $\times$ 10$^{21}$ cm$^{-2}$, of $\sim$2 $\times$ 10$^{5}$ $M_{\odot}$ assuming that H{\sc i} column density is uniform in the same area. Part of them formed the stars until now. If we take a ratio of the stars and the H{\sc i} gas, the star formation efficiency is estimated to be less than 10$^{-3}$. Normally, we can estimate the typical time scale of the collision from a ratio of the displacement in the complementary distribution and velocity separation, and we obtain $\sim$60 pc / 40 km s$^{-1}$ = 1.5 Myr. By considering the possible time span due to the cloud configuration, a large time span of a few Myr or longer may be applicable. 

The H{\sc i} is ionized around N83A, showing a small cavity of $\sim$15 pc in diameter (Figure \ref{fig7}). The cavity is seen in a velocity range at least 10.0 km s$^{-1}$ in the red cloud. The ALMA data shows 29 CO clouds surrounding N83 and N84. The total mass of the CO clouds amounts to $\sim$10$^{5}$ $M_{\odot}$. The CO clouds were possibly formed in the compressed H{\sc i} gas in a time scale of a few Myr. Numerical simulations of the colliding H{\sc i} flows by \citet{2012ApJ...759...35I} show the formation of CO clouds, which is demonstrated by the synthetic observations of the simulations \citep{2018ApJ...860...33F,2018arXiv181102224T}. The typical compressional time scale in the collision is estimated to be a ratio of the cluster size and the relative cloud velocity, 100 pc / 40 km s$^{-1}$ $\sim$ 2.5 Myr, which is roughly consistent with the youngest stellar ages above. The collision continued in a duration of at least a few Myr to trigger star formation with different ages from 2.3 Myr to $\sim$10 Myr due to the different path lengths between the two colliding components over 500 pc. It is likely that the H{\sc i} colliding flows eventually lead to the cluster formation as shown by the recent MHD simulations (R. Maeda et al. 2020, in preparation).

The collision direction is nearly in the north-south, where the red cloud and the blue cloud are approaching. The relative motion has the same sense in the both regions of NGC~602 and N83/N84, and the displacements are 147 pc in NGC~602 and 57 pc in N83/N84. Outside the SMC, it has been puzzling why only the NGC~602 and the N83/N84 regions are very active in high-mass star formation in spite of the low density environments. The displacements derived in the NGC~602 and the N83/N84 regions are both in the north to south direction with a scale of $\sim$60--150 pc, and are possibly interpreted in terms of a large kpc scale systematic gas motion driven by the tidal interaction between the LMC and the SMC. A possibility is that the red-shifted H{\sc i} gas is moving from the SMC toward the Magellanic Bridge and the star formation became active at the two colliding spots. This requires to be tested by three-dimensional numerical simulation of gas hydrodynamics. As suggested by \citet{2020arXiv200513750F}, the red-shifted cloud possibly is flowing toward the Magellanic Bridge, which was formed by the close encounter between the SMC and the LMC 200 Myr ago \citep{1990PASJ...42..505F,2007MNRAS.381L..11M}. \citet{2017PASJ...69L...5F} and \citet{2019ApJ...871...44T} showed that the encounter caused the two velocity components in the LMC, which are colliding with each other to form R136 and LHA 120-N 44 (N44). It is possible that the present collision is a counterpart in the SMC as suggested by \citet{2007MNRAS.381L..11M}, and suggests that the encounter has a broad impact on the high-mass star formation in the Magellanic system in particular in the southeast wing having low column density, where the present collision provides a viable explanation on the exceptionally active star formation.

\section{Conclusions} \label{sec:conc}
We have analyzed the new GASKAP H{\sc i} data at 30\arcsec~resolution, and discovered signatures of colliding H{\sc i} flows toward N83, N84 and nearby ten O-type stars. The main findings of the present work are summarized as follows;
\begin{enumerate}
\item The southeast region of the SMC where N83 and N84 are located have two velocity components with 35.1 km s$^{-1}$ velocity separation. The masses of the clouds are $\sim$(1--1.5) $\times$ 10$^{7}$ $M_{\odot}$ with typical column density of $\sim$3 $\times$ 10$^{21}$ cm$^{-2}$. The two components show complementary distribution with each other. With a displacement of 57 pc in the same sense with that found in the NGC~602 region. In addition, the two components are connected by bridge features in velocity.
\item These signatures in (1) suggest that the two components collided with each other over a time scale of $\sim$2 Myr. The collision continued in a duration of at least a few Myr to trigger star formation with different ages from 2.3 Myr to $\sim$10 Myr due to the different path lengths between the two colliding components over 500 pc. The interaction formed 29 CO clouds within $\sim$100 pc in diameter by the collisional compression in the southern edge of the red cloud, which lead to the formation of N83, N84 and six O-type stars.
\item The H{\sc i} is ionized around N83A, showing a small cavity of $\sim$15 pc in diameter. The small cavity is seen in a velocity range at least 10 km s$^{-1}$ in the red cloud. The CO clouds were possibly formed in the compressed H{\sc i} gas in a time scale of a few Myr. The star formation efficiency is estimated to be less than 10$^{-3}$ from a ratio of the O-type star mass and the H{\sc i} mass within 100 pc in diameter.
\item The origin of the two H{\sc i} components is explained as due to the close encounter between the LMC and the SMC 200 Myrs ago, which was proposed by \citet{1990PASJ...42..505F}. It is likely that the present collision is a counterpart in the SMC as suggested by \citet{2007MNRAS.381L..11M}, and suggests that the encounter has a broad impact on the high-mass star formation in the Magellanic system following NGC~602. \citet{2017PASJ...69L...5F} and \citet{2019ApJ...871...44T} showed that the encounter caused the two velocity components in the LMC, which are colliding with each other to form R136 and N44.
\end{enumerate}

\begin{ack}
The authors are grateful to Kenji Bekki for his insightful comments on the tidal interaction in the Magellanic system, which helped to improve the manuscript. We also thank Enrico Di Teodoro and Naomi McClure-Griffiths for sharing the model velocity field of the SMC with us. The Australian SKA Pathfinder is part of the Australia Telescope National Facility which is managed by CSIRO. Operation of ASKAP is funded by the Australian Government with support from the National Collaborative Research Infrastructure Strategy. ASKAP uses the resources of the Pawsey Supercomputing Centre. Establishment of ASKAP, the Murchison Radio-astronomy Observatory and the Pawsey Super- computing Centre are initiatives of the Australian Government, with support from the Government of Western Australia and the Science and Industry Endowment Fund. We acknowledge the Wajarri Yamatji people as the traditional owners of the Observatory site. This paper makes use of the following ALMA data: ADS/JAO.ALMA\#2018.1.01319.S. ALMA is a partnership of ESO (representing its member states), NSF (USA) and NINS (Japan), together with NRC (Canada), MOST and ASIAA (Taiwan), and KASI (Republic of Korea), in cooperation with the Republic of Chile. The Joint ALMA Observatory is operated by ESO, AUI/NRAO and NAOJ. In addition, publications from NA authors must include the standard NRAO acknowledgement: The National Radio Astronomy Observatory is a facility of the National Science Foundation operated under cooperative agreement by Associated Universities, Inc. Based on observations made with the NASA/ESA Hubble Space Telescope, and obtained from the Hubble Legacy Archive, which is a collaboration between the Space Telescope Science Institute (STScI/NASA), the Space Telescope European Coordinating Facility (ST-ECF/ESA) and the Canadian Astronomy Data Centre (CADC/NRC/CSA). This work was financially supported in part by JSPS KAKENHI Grant Numbers 15H05694, 19K14758, and 19H05075.
$Software:$ CASA (v 4.5.3.: \cite{2007ASPC..376..127M})
\end{ack}
\clearpage

\begin{figure}[ht]
\begin{center}
\includegraphics[width=100mm]{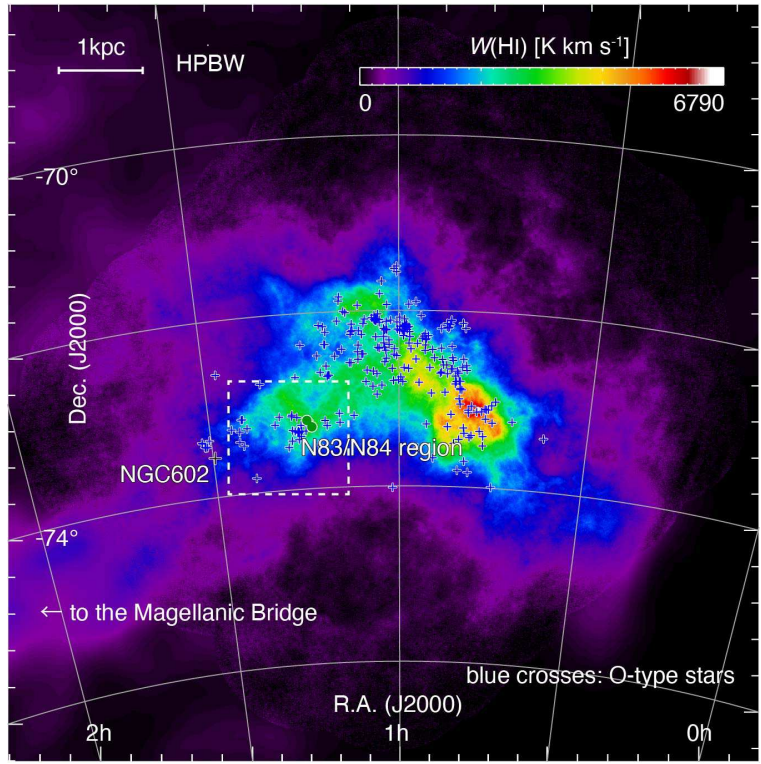}
\end{center}
\vspace*{0.5cm}
\caption{Integrated intensity map of ASKAP H{\sc i} whole the SMC \citep{2009IEEEP..97.1507D,2018NatAs...2..901M}. The integration velocity range is $V_{\rm {LSR}}$ = 50.4--253.6 km s$^{-1}$. The filled circles show the position of the N83/N84 region. The blue crosses show the positions of O-type stars. The beam size and scale bar are also shown in the top left corner. Focused region in the present paper is shown by the white dashed box.}
\label{fig1}
\end{figure}

\begin{figure}[ht]
\begin{center}
\includegraphics[width=130mm]{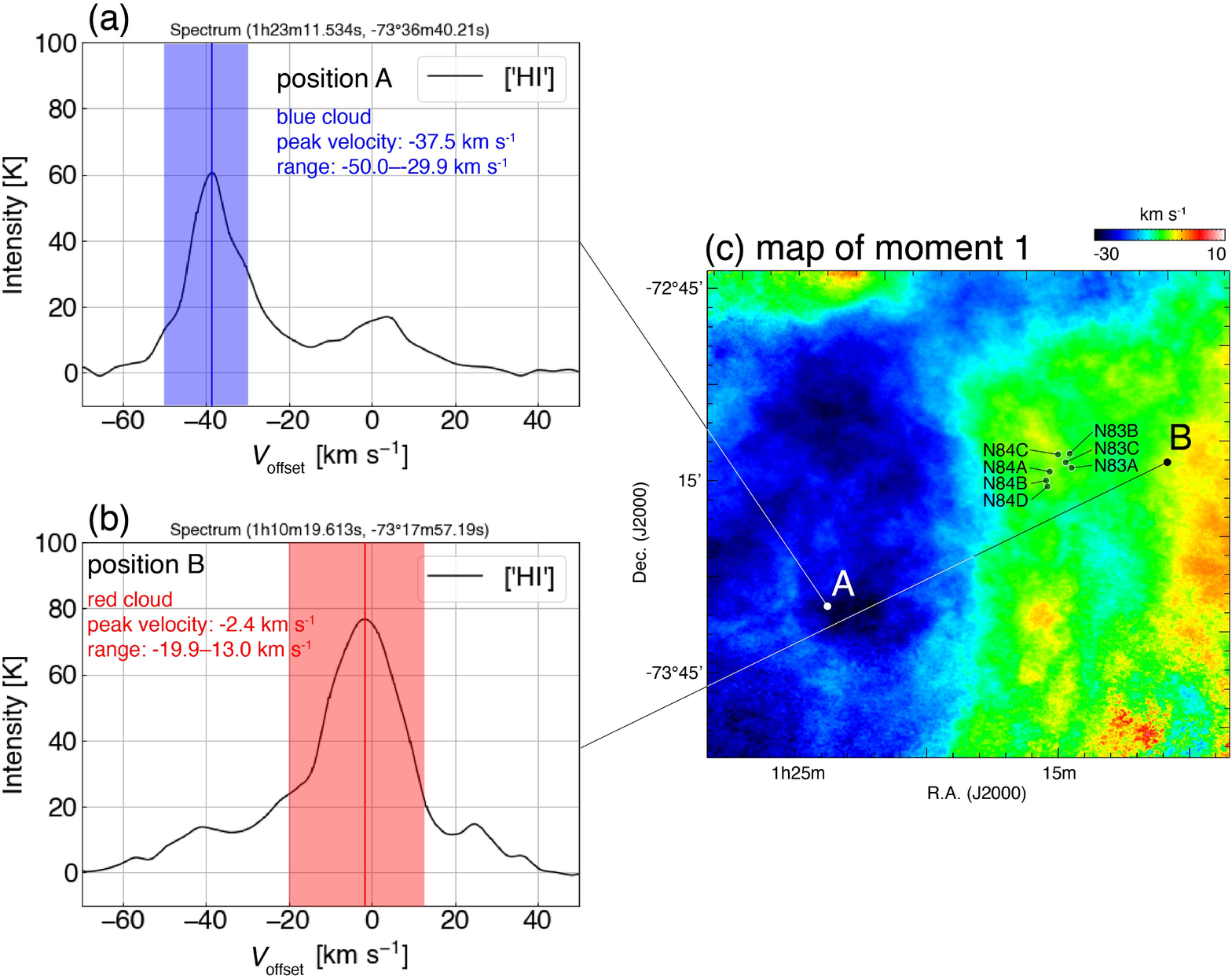}
\end{center}
\vspace*{0.5cm}
\caption{Typical H{\sc i} spectra at the positions of (a) A ($\alpha_\mathrm{J2000}$ = $1^{\mathrm{h}}23^{\mathrm{m}}11\fs5$, $\delta_\mathrm{J2000}$ = $-73{^\circ}36\arcmin40\farcs2$) and (b) B ($\alpha_\mathrm{J2000}$ = $1^{\mathrm{h}}10^{\mathrm{m}}19\fs6$, $\delta_\mathrm{J2000}$ = $-73{^\circ}17\arcmin57\farcs2$). The peak velocity and the velocity range are $-$37.5 km s$^{-1}$ and $-$50.0--$-$29.9 km s$^{-1}$ for the position A; and $-$2.4 km s$^{-1}$ and $-$19.9--13.0 km s$^{-1}$ for the position B, respectively. (c) Map of moment 1 toward the N83/N84 region derived using miriad task. The filled circles show the position of the N83/N84 region.}
\label{fig2}
\end{figure}

\begin{figure}[ht]
\begin{center}
\includegraphics[width=130mm]{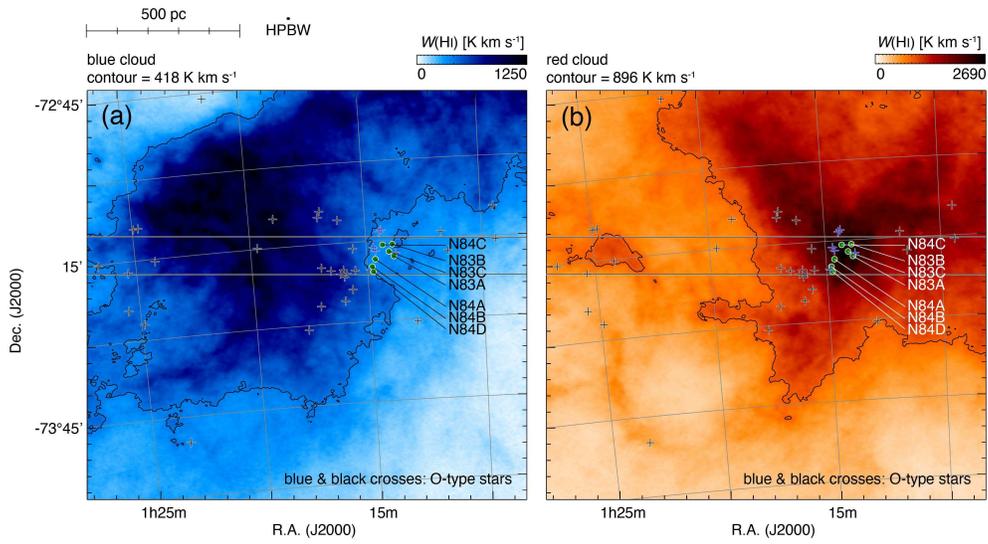}
\end{center}
\vspace*{0.5cm}
\caption{Intensity maps of (a) the blue cloud and (b) the red cloud. The integration velocity ranges of blue and red clouds are $V_{\rm {offset}}$ = $-$50.0--$-$29.9 km s$^{-1}$ and $V_{\rm {offset}}$ = $-$19.9--13.0 km s$^{-1}$, respectively. A contour of 20\% of the peak integrated intensity is shown in each panel. The filled circles are the same as those in Figure \ref{fig1}. The blue and black crosses represent positions of O-type stars discussed in the text and the other O-type stars, respectively. The black horizontal solid lines show the integration range of the right ascension-velocity diagram Figure \ref{fig4}. The beam size and scale bar are also shown in the top left corner.}
\label{fig3}
\end{figure}

\begin{figure}[ht]
\begin{center}
\includegraphics[width=120mm]{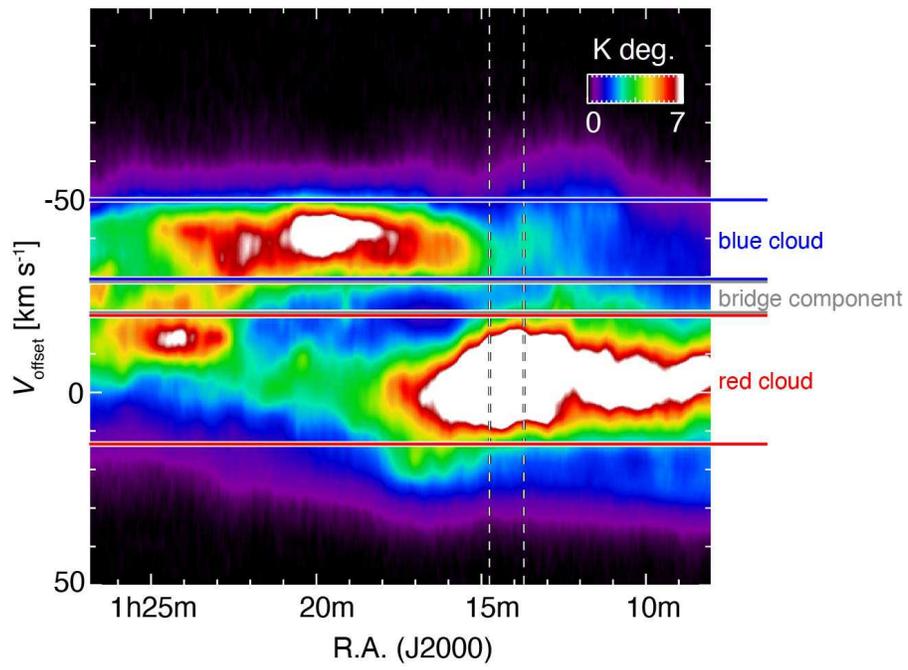}
\end{center}
\vspace*{0.5cm}
\caption{Right Ascension-velocity diagram of H{\sc i} toward the N83/N84 region. The integration range of Declination is from $-73{^\circ}16\arcmin48\farcs0$ to $-73{^\circ}09\arcmin36\farcs0$. The blue and red horizontal lines show the velocity ranges of blue and red clouds, respectively. The white dashed lines show the position of the N83/N84 region.}
\label{fig4}
\end{figure}

\begin{figure}[ht]
\begin{center}
\includegraphics[width=130mm]{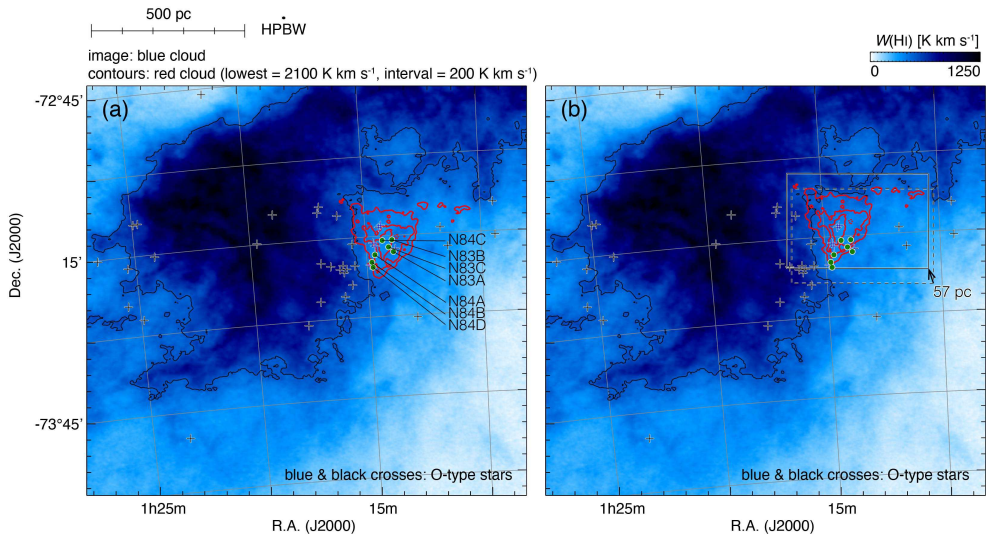}
\end{center}
\vspace*{0.5cm}
\caption{(a) Complementary spatial distribution of the blue cloud (image and black contour) and the red cloud (contours). The velocity ranges of the blue and red clouds are the same as shown in Figures \ref{fig3}(a) and \ref{fig3}(b), respectively. Contours of the red cloud are plotted every 200 K km s$^{-1}$ from 2100 K km s$^{-1}$. (b) Complementary spatial distribution of the blue cloud (image) and the red cloud (contours). The contours are spatially displaced 57 pc in the direction of north. The dashed and solid boxes show before and after displacement of the contours, respectively. The filled circles are the same as those in Figure \ref{fig1}. The blue and black crosses are the same as those in Figure \ref{fig3}. The beam size and scale bar are also shown in the top left corner.}
\label{fig5}
\end{figure}

\begin{figure}[ht]
\begin{center}
\includegraphics[width=100mm]{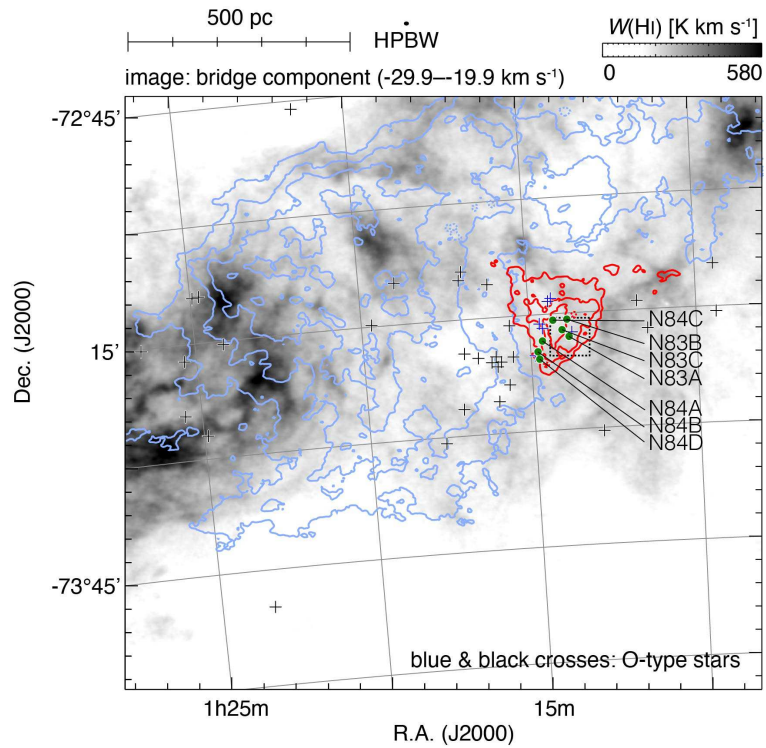}
\end{center}
\vspace*{0.5cm}
\caption{Distribution of the bridge component (gray scale) superposed on the contours of the blue cloud (light blue contours) and the red cloud (red contours). Contours of the blue cloud are plotted every 200 K km s$^{-1}$ from 530 K km s$^{-1}$ and of the red cloud are the same as those in Figure \ref{fig5}. The velocity ranges of the blue and red clouds are the same as shown in Figures \ref{fig3}(a) and \ref{fig3}(b), respectively. The velocity range of bridge component is $V_{\rm {offset}}$ = $-$29.9--$-$19.9 km s$^{-1}$. The filled circles are the same as those in Figure \ref{fig1}. The blue and black crosses are the same as those in Figure \ref{fig3}. Focused region in Figure \ref{fig7}(a) is shown by the black dashed box. The beam size and scale bar are also shown in the top left corner.}
\label{fig6}
\end{figure}

\begin{figure}[ht]
\begin{center}
\includegraphics[width=100mm]{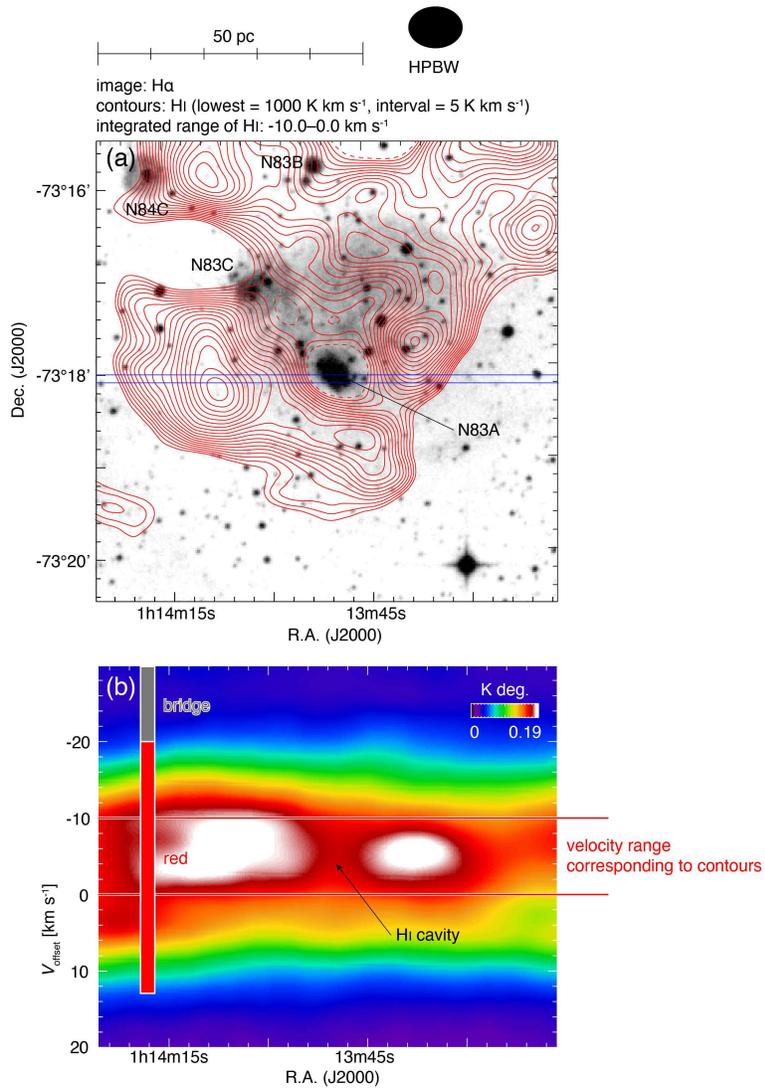}
\end{center}
\vspace*{0.5cm}
\caption{(a) H{\sc i} contour map integrated in the velocity range of $\it{V}$$_{\rm offset}$ = $-$10.0--0.0 km s$^{-1}$ superposed with the optical image toward N83A obtained from DSS2. Contours are plotted every 5 K km s$^{-1}$ from 1000 K km s$^{-1}$. The blue horizontal lines show the integration range of the right ascension-velocity diagram Figure 7(b). The beam size and scale bar are also shown in the top. (b) Right Ascension-velocity diagram of H{\sc i}. The integration range of Declination is from $-73{^\circ}18\arcmin04\farcs8$ to $-73{^\circ}17\arcmin59\farcs7$. The red horizontal solid lines show the H{\sc i} integration velocity range in Figure 7(a). The gray and red strips show the velocity ranges of the bridge component and red cloud, respectively.}
\label{fig7}
\end{figure}

\begin{figure}[ht]
\begin{center}
\includegraphics[width=130mm]{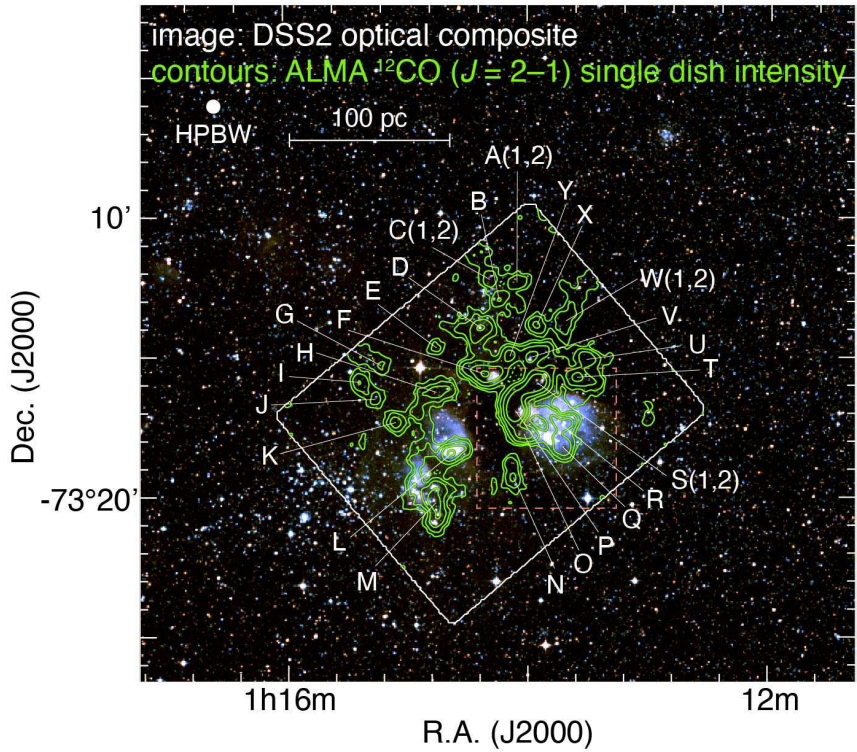}
\end{center}
\caption{Contour map of ALMA $^{12}$CO($J$ = 2--1) integrated intensity superposed with the DSS2 optical composite image toward the N83/N84 region. The integration velocity range of CO is $V_{\rm {LSR}}$ = 148.2--175.2 km s$^{-1}$. Contour levels of CO are 0.37 (5$\sigma$), 0.75, 1.50, 3.00, and 5.99 K km s$^{-1}$. The CO peaks A--Y discussed in {\S}\ref{sec:discuss2} are shown. Focused region in Figure \ref{fig7}(a) is shown by the red dashed box, and an observed area is shown with white solid box. The beam size and scale bar are also shown in the top left corner.}
\label{fig8}
\end{figure}

\begin{appendix}
\begin{center}
APPENDIX\\
Subtraction of the galactic rotation velocity component of the SMC
\end{center}
To subtract the galactic rotation velocity component of the SMC, we used the following method.
\begin{enumerate}
\item Galactic rotation velocity component\\
We used rotation velocity map of the SMC obtained from \citet{2019MNRAS.483..392D}. The H{\sc i} gas component of the SMC is modeled as a rotation disk on the assumption of non-negligible angular size, moving into the plane of the sky, and undergoing nutation/precession motions. For more details, see \citet{2019MNRAS.483..392D}.
\item Subtraction of the galactic rotation velocity component\\
We used the following equation to obtain radial velocity of the SMC ($V_{\rm {offset}}$),
\begin{equation}
V_{\rm {offset}} = V_{\rm {LSR}} - V_{\rm {rotation}},
\end{equation}
where $V_{\rm {rotation}}$ is the rotation velocity of the SMC. By using a liner interpolation algorithm, we increased the velocity resolution of the H{\sc i} data from 3.90 km s$^{-1}$ ch$^{-1}$ to 0.12 km s$^{-1}$ ch$^{-1}$ in order to apply the above equation. The velocity resolution of 0.12 km s$^{-1}$ ch$^{-1}$ is significantly smaller than the typical resolution of the rotation velocity of 0.16 km s$^{-1}$ toward the N83/N84 region. We then applied the above equation for each pixel of the region represented by the white dashed box in Figure \ref{fig1} using the following algorithm.
\item Algorithm of subtraction
\begin{enumerate}
\item Matching array sizes in the x (Right Ascension) and y (Declination) directions of the H{\sc i} data and the rotation velocity map using the MIRIAD software \citep{1995ASPC...77..433S}.
\item We define the rotation velocity at a position of (x$_{i}$, y$_{j}$) as $V_{\rm {rot,}\it{ij}}$. The suffix of $i$ and $j$ denotes the coordinates along the x and y axis of the rotation velocity map, respectively. We next extract the velocity array of the H{\sc i} data z$_{ij}$ at a position of (x$_{i}$, y$_{j}$), which refers to as H{\sc i}$_{\rm{z,}\it{ij}}$. We search a pixel in H{\sc i}$_{\rm{z,}\it{ij}}$ which shows the closest value of $V_{\rm {rot,}\it{ij}}$, and is defined as H{\sc i}$_{\rm{zrot,}\it{ij}}$. In addition, we define a reference pixel in the velocity array of the H{\sc i} data as H{\sc i}$_{\rm {zref}}$, and shift H{\sc i}$_{\rm{z,}\it{ij}}$ by $| \rm H\textsc{i}_{\rm {zrot,}\it{ij}} - \rm H\textsc{i}_{\rm {zref}} |$ in the velocity direction. Here we use a single value of H{\sc i}$_{\rm {zref}}$ by conducting this operation in all areas.
\item Finally, we subtract the rotation component by redefining H{\sc i}$_{\rm {zref}}$ as 0 km s$^{-1}$.
\end{enumerate}
\end{enumerate}
\end{appendix}
\clearpage

\begin{table}[ht]
\tbl{Physical properties of H{\sc i} clouds}{%
\begin{tabular}{lccccc} \hline \hline
Cloud name & Velocity range & Peak velocity & Peak column density & Mass & Cloud radius \\
& [km s$^{-1}$] & [km s$^{-1}$] & [$\times10^{21}$ cm$^{-2}$] & [$\times10^{7} M_{\odot}$] & [pc] \\
(1) & (2) & (3) & (4) & (5) & (6) \\ \hline
Blue cloud & $-$50.0--$-$29.9 & $-$37.5 & 2.3 & 1.0\phantom{0} & 600\\
Bridge component & $-$29.9--$-$19.9 & ----- & 1.2 & 0.25 & 400\\
Red cloud & $-$19.9--13.0 & $-$2.4 & 4.9 & 1.5\phantom{0} & 500\\ \hline
\end{tabular}}
\begin{tabnote}
\hangindent6pt\noindent
Note. --- Col. (1): Cloud name. Col. (2) Velocity range defined by visual inspection. Col. (3): Peak velocity derived from a Gaussian fitting. Col. (4): Peak column density derived by using equation (2). Col. (5): Mass derived from column density. Col. (6): Cloud radius defined as an effective radius $R_{\rm {HI}}$ derived by using equation (3).
\end{tabnote}
\label{t1}
\end{table}
\clearpage

\begin{table}[ht]
\tbl{Physical properties of CO clouds}{%
\begin{tabular}{lcccccccc} \hline \hline
Object & R.A. & Dec. & $T_{\rm {peak}}$ & $V_{\rm {peak}}$ & $\Delta V$ & $D_{\rm {CO}}$ & $M_{\rm {CO}}$ & remark \\ 
& [hms] & [dms] & [K] & [km s$^{-1}$] & [km s$^{-1}$] & [pc] & [$M_{\odot}$] & \\
(1) & (2) & (3) & (4) & (5) & (6) & (7) & (8) & \\ \hline
A1 & $1^{\mathrm{h}}14^{\mathrm{m}}07\fs5$ & $-73{^\circ}12\arcmin21\farcs2$ & 0.48 & 163.9 & 1.5 & 10.2 & \phantom{0}\phantom{0}800 & -- \\
A2 & $1^{\mathrm{h}}14^{\mathrm{m}}03\fs9$ & $-73{^\circ}12\arcmin30\farcs2$ & 0.28 & 168.2 & 2.9 & \phantom{0}9.8 & \phantom{0}\phantom{0}900 & -- \\
B & $1^{\mathrm{h}}14^{\mathrm{m}}14\fs3$ & $-73{^\circ}12\arcmin59\farcs4$ & 0.69 & 164.3 & 1.7 & \phantom{0}9.6 & \phantom{0}1200 & -- \\
C1 & $1^{\mathrm{h}}14^{\mathrm{m}}19\fs4$ & $-73{^\circ}12\arcmin18\farcs8$ & 0.48 & 166.7 & 1.9 & 10.3 & \phantom{0}1100 & -- \\
C2 & $1^{\mathrm{h}}14^{\mathrm{m}}18\fs4$ & $-73{^\circ}12\arcmin09\farcs9$ & 0.17 & 161.0 & 1.0 & \phantom{0}9.0 & \phantom{0}\phantom{0}200 & -- \\
D & $1^{\mathrm{h}}14^{\mathrm{m}}22\fs6$ & $-73{^\circ}14\arcmin00\farcs1$ & 1.05 & 163.6 & 2.6 & 10.7 & \phantom{0}3400 & -- \\
E & $1^{\mathrm{h}}14^{\mathrm{m}}46\fs6$ & $-73{^\circ}14\arcmin40\farcs2$ & 0.52 & 161.9 & 1.7 & \phantom{0}8.5 & \phantom{0}\phantom{0}700 & -- \\
F & $1^{\mathrm{h}}14^{\mathrm{m}}23\fs7$ & $-73{^\circ}15\arcmin39\farcs1$ & 1.71 & 160.4 & 2.6 &11.6 & \phantom{0}7500 & -- \\
G & $1^{\mathrm{h}}15^{\mathrm{m}}11\fs6$ & $-73{^\circ}15\arcmin20\farcs1$ & 0.58 & 164.0 & 1.3 & 10.1 & \phantom{0}\phantom{0}800 & -- \\
H & $1^{\mathrm{h}}14^{\mathrm{m}}39\fs9$ & $-73{^\circ}16\arcmin14\farcs8$ & 0.89 & 165.8 & 1.5 & 12.1 & \phantom{0}2200 & -- \\
I & $1^{\mathrm{h}}15^{\mathrm{m}}23\fs1$ & $-73{^\circ}16\arcmin02\farcs6$ & 0.80 & 166.9 & 1.6 & 10.1 & \phantom{0}1500 & -- \\
J & $1^{\mathrm{h}}15^{\mathrm{m}}15\fs9$ & $-73{^\circ}16\arcmin34\farcs3$ & 0.38 & 168.3 & 2.0 & 10.7 & \phantom{0}1000 & -- \\
K & $1^{\mathrm{h}}15^{\mathrm{m}}05\fs5$ & $-73{^\circ}17\arcmin28\farcs5$ & 0.68 & 166.6 & 2.9 & \phantom{0}9.3 & \phantom{0}1900 & -- \\
L & $1^{\mathrm{h}}14^{\mathrm{m}}38\fs5$ & $-73{^\circ}18\arcmin38\farcs8$ & 0.80 / 0.26 & 167.1 / 170.1 & 2.3 / 2.7 & \phantom{0}9.3 & \phantom{0}2700 & double peak \\
M & $1^{\mathrm{h}}14^{\mathrm{m}}46\fs9$ & $-73{^\circ}20\arcmin04\farcs2$ & 1.76 & 167.9 & 1.9 & 14.7 & \phantom{0}8600 & -- \\
N & $1^{\mathrm{h}}14^{\mathrm{m}}07\fs7$ & $-73{^\circ}19\arcmin22\farcs0$ & 0.71 & 166.4 & 2.2 & 12.4 & \phantom{0}2700 & -- \\
O & $1^{\mathrm{h}}14^{\mathrm{m}}02\fs4$ & $-73{^\circ}17\arcmin49\farcs7$ & 1.38 & 162.9 & 2.6 & 10.8 & \phantom{0}4700 & -- \\
P & $1^{\mathrm{h}}14^{\mathrm{m}}07\fs1$ & $-73{^\circ}17\arcmin00\farcs2$ & 4.51 & 162.0 & 2.8 & 11.1 & 17700 & -- \\
Q & $1^{\mathrm{h}}13^{\mathrm{m}}42\fs1$ & $-73{^\circ}18\arcmin16\farcs8$ & 0.70 & 158.5 & 3.3 & 10.1 & \phantom{0}2400 & -- \\
R & $1^{\mathrm{h}}13^{\mathrm{m}}47\fs8$ & $-73{^\circ}17\arcmin13\farcs8$ & 0.65 & 156.2 & 1.9 & 12.1 & \phantom{0}1900 & -- \\
S1 & $1^{\mathrm{h}}13^{\mathrm{m}}58\fs7$ & $-73{^\circ}16\arcmin08\farcs5$ & 3.10 & 161.7 & 2.5 & 11.4 & 13100 & -- \\
S2 & $1^{\mathrm{h}}13^{\mathrm{m}}56\fs6$ & $-73{^\circ}15\arcmin43\farcs8$ & 1.07 & 151.3 & 2.2 & \phantom{0}9.7 & \phantom{0}2500 & -- \\
T & $1^{\mathrm{h}}13^{\mathrm{m}}35\fs3$ & $-73{^\circ}15\arcmin46\farcs0$ & 0.93 / 0.69 & 162.7 / 165.2 & 2.2 / 1.6 & 10.9 & \phantom{0}4100 & double peak \\
U & $1^{\mathrm{h}}13^{\mathrm{m}}30\fs1$ & $-73{^\circ}15\arcmin10\farcs0$ & 1.50 & 159.2 & 1.7 & \phantom{0}9.3 & \phantom{0}2500 & -- \\
V & $1^{\mathrm{h}}13^{\mathrm{m}}55\fs1$ & $-73{^\circ}14\arcmin54\farcs3$ & 1.41 & 151.5 & 1.7 & 12.6 & \phantom{0}4700 & -- \\
W1 & $1^{\mathrm{h}}13^{\mathrm{m}}41\fs5$ & $-73{^\circ}13\arcmin51\farcs3$ & 0.55 & 160.7 & 1.8 & 12.1 & \phantom{0}1600 & -- \\
W2 & $1^{\mathrm{h}}13^{\mathrm{m}}45\fs2$ & $-73{^\circ}13\arcmin44\farcs5$ & 0.22 & 156.5 & 2.0 & 10.4 & \phantom{0}\phantom{0}600 & -- \\
X & $1^{\mathrm{h}}13^{\mathrm{m}}55\fs6$ & $-73{^\circ}13\arcmin55\farcs8$ & 0.82 & 168.0 & 1.9 & \phantom{0}8.7 & \phantom{0}1300 & -- \\
Y & $1^{\mathrm{h}}14^{\mathrm{m}}09\fs1$ & $-73{^\circ}14\arcmin58\farcs7$ & 1.65 & 158.3 & 2.2 & \phantom{0}9.5 & \phantom{0}3700 & -- \\ \hline
\noalign{\vskip3pt} 
\end{tabular}}
\begin{tabnote}
\hangindent6pt\noindent
Note. --- Col. (1): Cloud name. Cols. (2)--(3): Position of the observed point with the maximum integrated intensity of $^{12}$CO($J$ = 2--1). Cols. (4)--(6): Observed properties of the $^{12}$CO($J$ = 2--1) spectra obtained at the peak positions of the CO clouds. Col. (4): Peak radiation temperature $T_{\rm {peak}}$. Col. (5): Center velocity $V_{\rm {peak}}$ ($V_{\rm {LSR}}$) derived from single or double Gaussian fitting. Col. (6): FWHM line width $\Delta V$. Col. (7): Cloud size derived by using equation (6). Col. (8): Mass of the clouds derived by using the relationship between the molecular hydrogen column density and the $^{12}$CO($J$ = 1--0) intensity using equations (4) and (5). Here we assume the intensity ratio of $^{12}$CO($J$ = 2--1)/$^{12}$CO($J$ = 1--0) = 0.9 \citep{2003ApJ...595..167B}.
\end{tabnote}
\label{t2}
\end{table}
\clearpage

\clearpage

\clearpage

\clearpage

\clearpage

\clearpage

\clearpage

\clearpage

\clearpage


\begin{thebibliography}{}

\bibitem[Bekki \& Chiba(2007)]{2007PASA...24...21B} Bekki, K., \& Chiba, M.\ 2007, PASA, 24, 21
\bibitem[Bolatto et al.(2003)]{2003ApJ...595..167B} Bolatto, A.~D., Leroy, A., Israel, F.~P., et al.\ 2003, \apj, 595, 167
\bibitem[Bonanos et al.(2010)]{2010AJ....140..416B} Bonanos, A.~Z., Lennon, D.~J., K{\"o}hlinger, F., et al.\ 2010, \aj, 140, 416
\bibitem[Bratsolis et al.(2004)]{2004A&A...423..919B} Bratsolis, E., Kontizas, M., \& Bellas-Velidis, I.\ 2004, \aap, 423, 919
\bibitem[DeBoer et al.(2009)]{2009IEEEP..97.1507D} DeBoer, D.~R., Gough, R.~G., Bunton, J.~D., et al.\ 2009, IEEE Proceedings, 97, 1507
\bibitem[Di Teodoro et al.(2019)]{2019MNRAS.483..392D} Di Teodoro, E.~M., McClure-Griffiths, N.~M., Jameson, K.~E., et al.\ 2019, \mnras, 483, 392
\bibitem[Finn et al.(2019)]{2019ApJ...874..120F} Finn, M.~K., Johnson, K.~E., Brogan, C.~L., et al.\ 2019, \apj, 874, 120
\bibitem[Fujimoto \& Noguchi(1990)]{1990PASJ...42..505F} Fujimoto, M., \& Noguchi, M.\ 1990, \pasj, 42, 505
\bibitem[Fujita et al.(2020)]{2020arXiv200313925F} Fujita, S., Sano, H., Enokiya, R., et al.\ 2020, arXiv e-prints, arXiv:2003.13925
\bibitem[Fukui et al.(2017)]{2017PASJ...69L...5F} Fukui, Y., Tsuge, K., Sano, H., et al.\ 2017, \pasj, 69, L5
\bibitem[Fukui et al.(2018)]{2018ApJ...860...33F} Fukui, Y., Hayakawa, T., Inoue, T., et al.\ 2018, \apj, 860, 33
\bibitem[Fukui et al.(2020)]{2020arXiv200513750F} Fukui, Y., Ohno, T., Tsuge, K., et al.\ 2020, arXiv e-prints, arXiv:2005.13750
\bibitem[Genzel et al.(1998)]{1998Natur.395..859G} Genzel, R., Lutz, D., \& Tacconi, L.\ 1998, \nat, 395, 859
\bibitem[Henize(1956)]{1956ApJS....2..315H} Henize, K.~G.\ 1956, \apjs, 2, 315
\bibitem[Inoue \& Inutsuka(2012)]{2012ApJ...759...35I} Inoue, T., \& Inutsuka, S.-i.\ 2012, \apj, 759, 35
\bibitem[Lortet \& Testor(1988)]{1988A&A...194...11L} Lortet, M.-C., \& Testor, G.\ 1988, \aap, 194, 11
\bibitem[Luks \& Rohlfs(1992)]{1992A&A...263...41L} Luks, T., \& Rohlfs, K.\ 1992, \aap, 263, 41
\bibitem[McClure-Griffiths et al.(2018)]{2018NatAs...2..901M} McClure-Griffiths, N.~M., D{\'e}nes, H., Dickey, J.~M., et al.\ 2018, Nature Astronomy, 2, 901
\bibitem[McMullin et al.(2007)]{2007ASPC..376..127M} McMullin, J.~P., Waters, B., Schiebel, D., et al.\ 2007, Astronomical Data Analysis Software and Systems XVI, 127
\bibitem[Mizuno et al.(2001)]{2001PASJ...53L..45M} Mizuno, N., Rubio, M., Mizuno, A., et al.\ 2001, \pasj, 53, L45
\bibitem[Muller \& Bekki(2007)]{2007MNRAS.381L..11M} Muller, E., \& Bekki, K.\ 2007, \mnras, 381, L11
\bibitem[Muraoka et al.(2017)]{2017ApJ...844...98M} Muraoka, K., Homma, A., Onishi, T., et al.\ 2017, \apj, 844, 98
\bibitem[Rohlfs \& Luks(1991)]{1991IAUS..148...63R} Rohlfs, K., \& Luks, T.\ 1991, in The Magellanic Clouds, ed. R. Haynes \& D. Milne (Dordrecht: Kluwer Academic Publishers), 63
\bibitem[Rohlfs \& Luks(1993)]{1993LNP...416..115R} Rohlfs, K., \& Luks, T.\ 1993, in New Aspects of Magellanic Cloud Research, ed B. Baschek, G. Klare, \& J. Lequeux (Berlin: Springer-Verlag), 115
\bibitem[Sault et al.(1995)]{1995ASPC...77..433S} Sault, R.~J., Teuben, P.~J., \& Wright, M.~C.~H.\ 1995, Astronomical Data Analysis Software and Systems IV, 433
\bibitem[Stanimirovi{\'c} et al.(2004)]{2004ApJ...604..176S} Stanimirovi{\'c}, S., Staveley-Smith, L., \& Jones, P.~A.\ 2004, \apj, 604, 176
\bibitem[Tachihara et al.(2018)]{2018arXiv181102224T} Tachihara, K., Fukui, Y., Hayakawa, T., et al.\ 2018, arXiv e-prints, arXiv:1811.02224
\bibitem[Testor \& Lortet(1987)]{1987A&A...178...25T} Testor, G., \& Lortet, M.-C.\ 1987, \aap, 178, 25
\bibitem[Tsuge et al.(2019a)]{2019ApJ...871...44T} Tsuge, K., Sano, H., Tachihara, K., et al.\ 2019a, \apj, 871, 44
\bibitem[Tsuge et al.(2019b)]{2019arXiv190905240T} Tsuge, K., Fukui, Y., Tachihara, K., et al.\ 2019b, arXiv e-prints, arXiv:1909.05240
\bibitem[Tsuge et al.(2020)]{2020arXiv200504075T} Tsuge, K., Tachihara, K., Fukui, Y., et al.\ 2020, arXiv e-prints, arXiv:2005.04075
\bibitem[Weaver et al.(1977)]{1977ApJ...218..377W} Weaver, R., McCray, R., Castor, J., et al.\ 1977, \apj, 218, 377
\bibitem[Whitmore et al.(2014)]{2014ApJ...795..156W} Whitmore, B.~C., Brogan, C., Chandar, R., et al.\ 2014, \apj, 795, 156
\bibitem[Wilson et al.(2000)]{2000ApJ...542..120W} Wilson, C.~D., Scoville, N., Madden, S.~C., et al.\ 2000, \apj, 542, 120

\end{thebibliography}
\end{document}